\documentclass{appolb}

\newcommand{\KV}{{\mbox{$\kappa \sigma^{2}$}}}
\newcommand{\SD}{{\mbox{$S \sigma$}}}

\newcommand{\axi}{$\overline{\Xi}^+$}
\newcommand{\xim}{$\Xi^-$}
\newcommand{\alam}{$\overline{\Lambda}$}
\newcommand{\lam}{$\Lambda$}

\usepackage{epsfig}

\begin{document}
\title{Highlights from the STAR experiment at RHIC%
\thanks{Presented at the International Conference on Strangeness in Quark Matter, 18-24 September 2011, Krakow, Poland}%
}
\author{Sonia Kabana (for the STAR Collaboration)
\address{Laboratoire de physique subatomique et des technologies associees, (SUBATECH), 4 rue Alfred Kastler,
 44307 Nantes, France}
}
\maketitle
\begin{abstract}
\noindent
Experiments using heavy ion collisions at ultrarelativistic energies aim  
 to  explore the QCD phase transition and map out the QCD phase diagram.
A wealth of remarkable results in this field
 have been reported recently,
for example  the $\Upsilon$  suppression discovered recently.
We discuss recent results from the STAR experiment 
focusing on strangeness, charm and beauty production.
\end{abstract}

\section{Introduction}
Experiments studying heavy ion collisions at ultrarelativistic energies
aim to reproduce and study in the laboratory one of the anticipated phase transitions of the early universe,
namely the QCD phase transition between partons and hadrons
predicted by lattice QCD
 at a critical temperature of ~160-180 MeV 
 and to map out the QCD phase diagram.

The STAR experiment at RHIC 
with the help of recent important upgrades,
including a full-barrel Time-of-Flight (TOF) detector allowing for better particle identification, an
upgraded DAQ and implementation of a High Level Trigger (HLT) system, has been able to
accumulate  a vast amount of high quality data from $p+p$ and Au+Au collisions.
In addition STAR has taken data of Au+Au collisions performing a low
energy scan covering the $\sqrt{s_{NN}}$ of 7.7 to 39 GeV, aiming to discover a possible
critical point, to study with precision the phase transition characteristics, and map
out the phase diagram of QCD.
In this paper, we present highlights from recent results of the STAR experiment, 
with a main focus on strangeness, charm and beauty production.

\section{Charm and Beauty}

Open beauty ($B$) and charm ($C$) are prominent probes of the sQGP medium produced in central 
ultrarelativistic heavy ion collisions.
STAR is the only experiment at RHIC able to measure charm through direct charmed particle
identification in addition to the measurement of charm and beauty through electrons originated from
their decays.
Electrons are identified using the $dE/dx$ information measured with the Time Projection Chamber (TPC),
the information of the electromagnetic calorimeter at high $p_T$, and using the new TOF detector
at low $p_T$. These detectors cover the midrapidity region and full azimuthal angle.

\begin{figure}
\begin{center}
\begin{tabular}{c}
\\
\includegraphics[height=1.9in,width=2.23in]{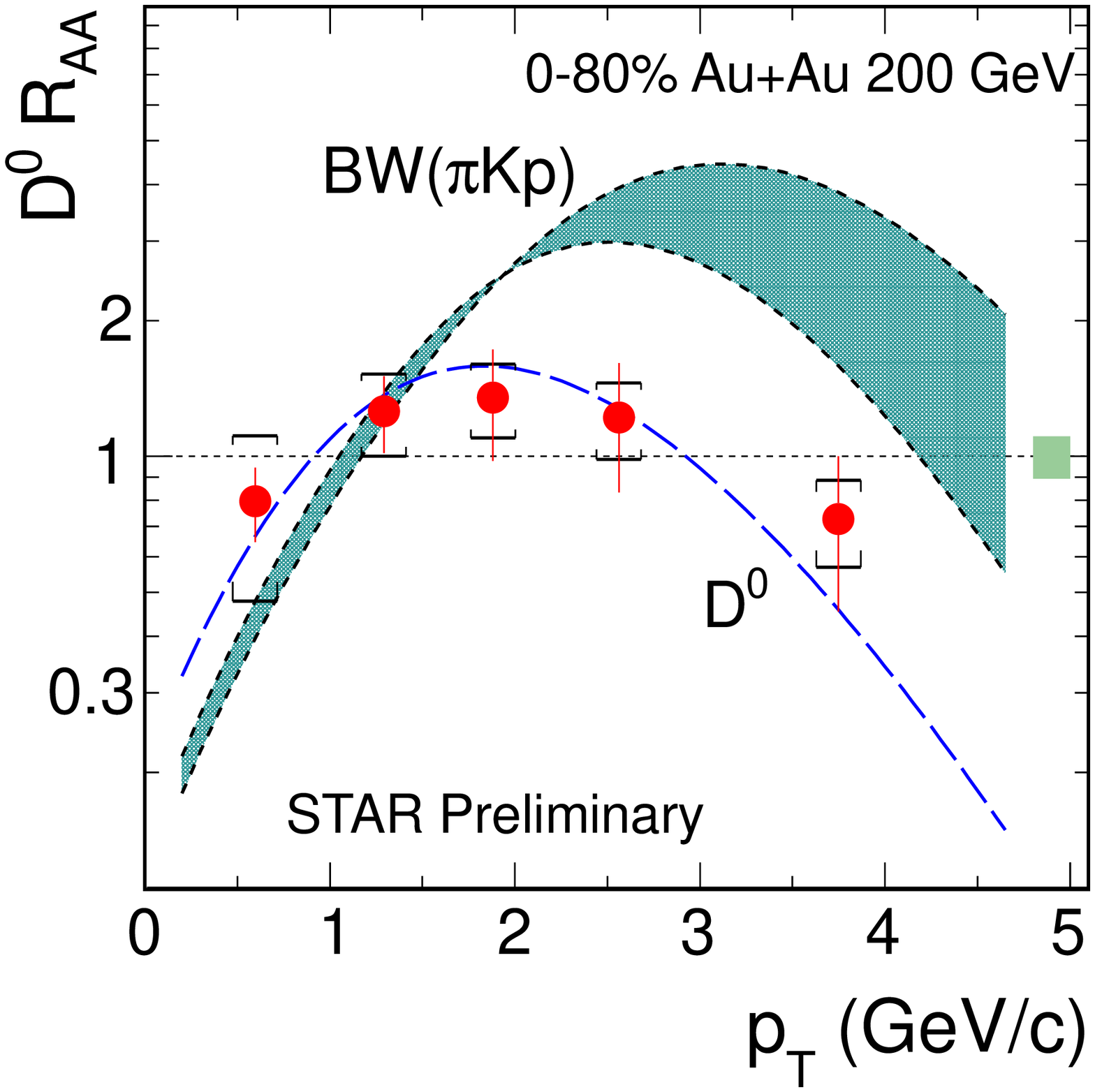}
\includegraphics[height=1.9in,width=2.23in]{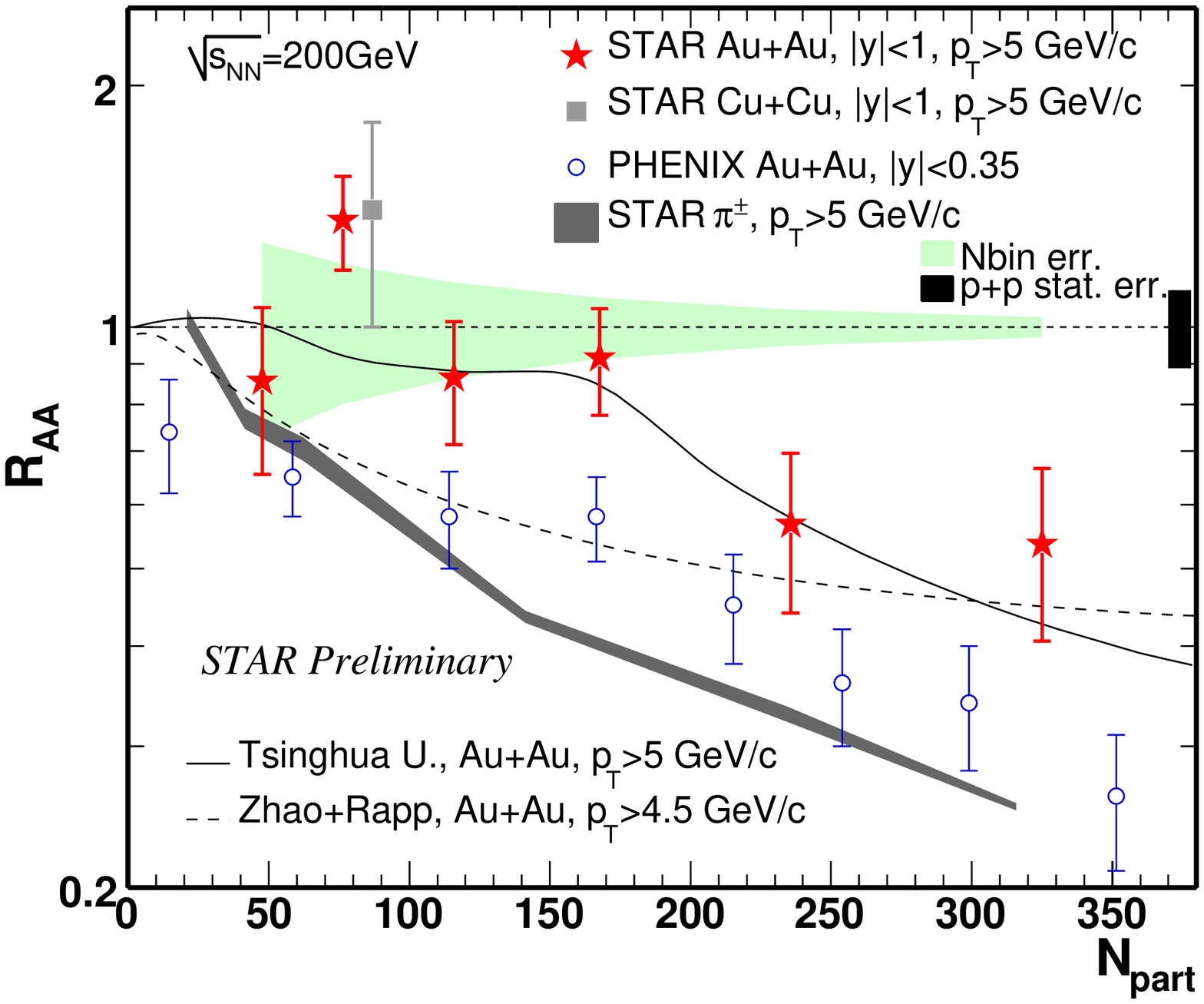}
\end{tabular}
\end{center}
\caption{
Left Panel: Fully reconstructed $D^0$ nuclear modification factor $R_{AA}$ as a function of $p_T$
in Au+Au collisions at $\sqrt{s_{NN}}$=200 GeV.
The dashed curve shows a blast-wave fit \cite{yifei_zhang_qm2011}. 
The shaded band is the predicted $D^0$ $R_{AA}$ with blast-wave parameters from light-quark hadrons.
Right Panel: $R_{AA}$ vs $N_{part}$ for  $J/\psi$ at high $p_T>$ 5 GeV/c (stars for Au+Au
 and one grey rectangle for Cu+Cu collisions) and
$J/\psi$ at low $p_T$=0-5 GeV/c in Au+Au collisions (open circles) measured by PHENIX, 
and
 high $p_T$ pions (dark thick line).
The solid and dashed thin lines show two theoretical calculations (see \cite{zebo_tang_qm2011} for details).
}
\label{fig1}
\end{figure}

The charm cross section at midrapidity has been
measured by STAR in a large variety of collisions 
($p+p$, $d$+Au, Cu+Cu, Au+Au)
and was found to scale with the number of binary collisions
  indicating production through initial hard scattering
  \cite{yifei_zhang_qm2011}. 
 Beauty and charm cross sections have been measured
 in $p+p$ collisions at 200 GeV and were found consistent with  
FONLL calculations 
\cite{NPE_pp_spectra}.
New results show that the beauty contribution to non-photonic electrons in $p+p$ collisions
 at 500 GeV is more than 60\% at $p_T$ higher than 8 GeV/c
\cite{wei_li_sqm2011}.

\begin{figure}
\begin{center}
\begin{tabular}{c}
\\
\includegraphics[height=1.9in,width=2.23in]{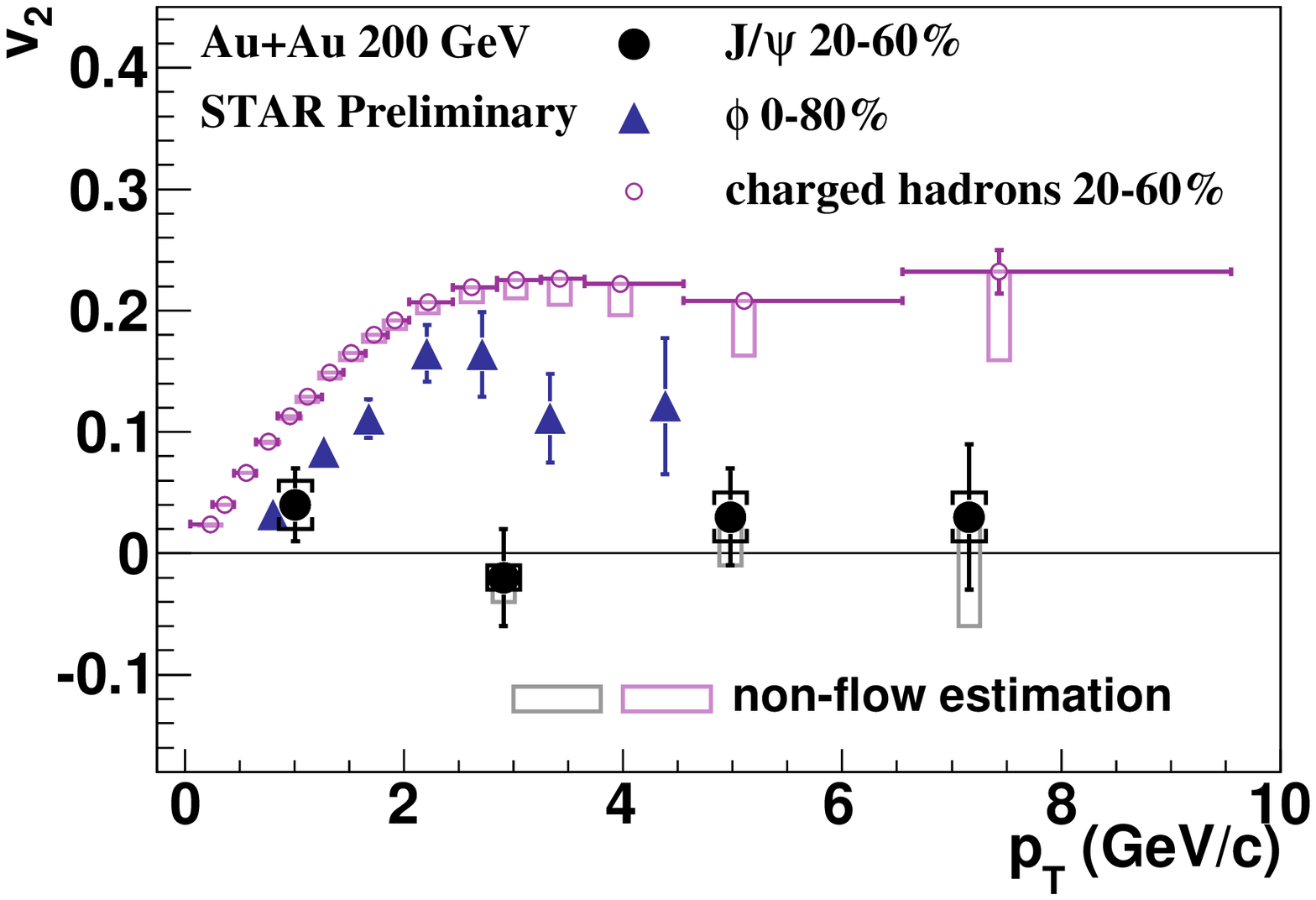}
\raisebox{-.05\height}
{
 \includegraphics[height=1.9in,width=2.33in]{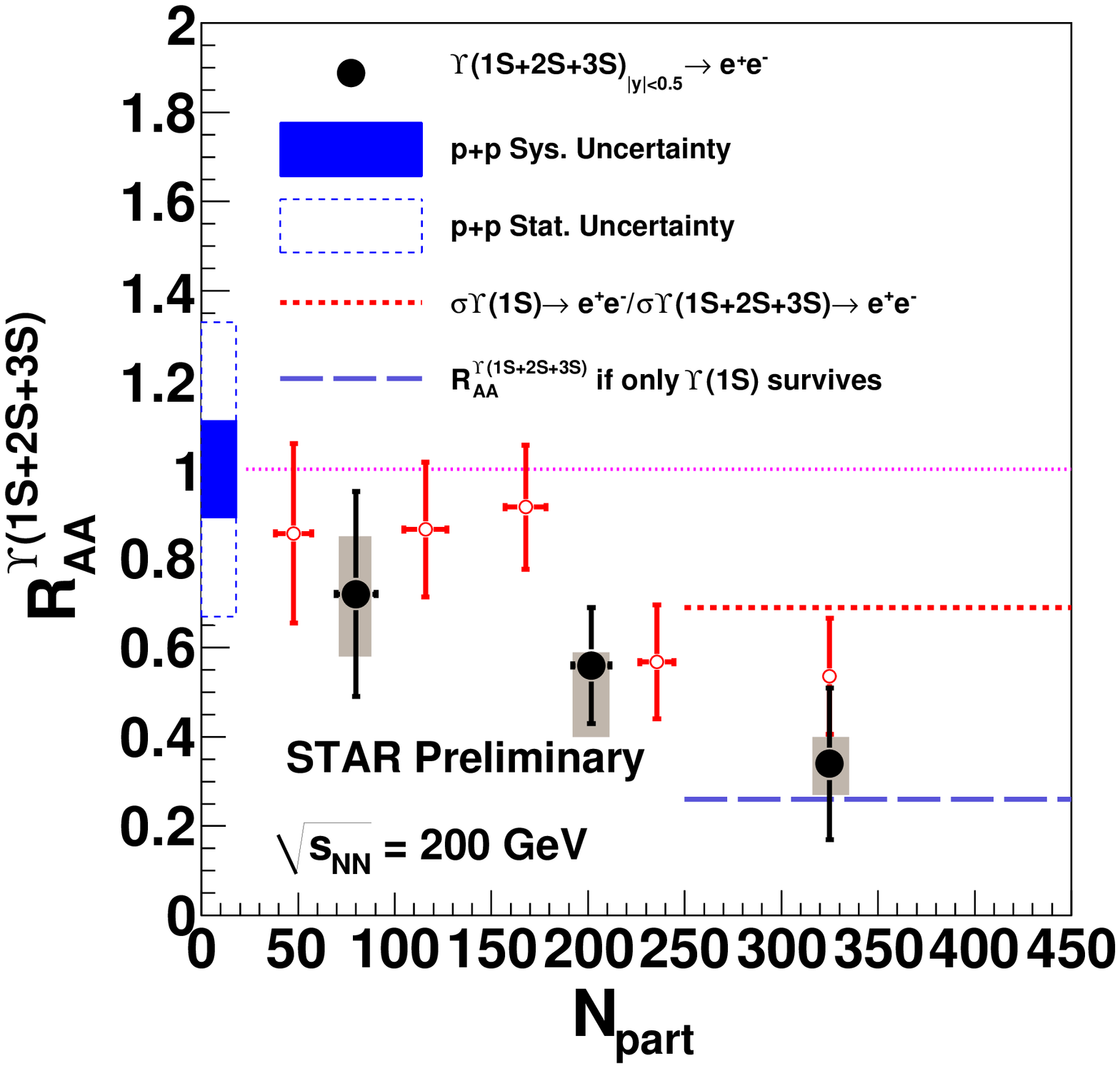}
}
\end{tabular}
\end{center}
 \caption{Left Panel:
$v_{2}$ as a function of $p_{T}$ for $J/\psi$ as well as charged hadrons and $\phi$ meson
in mid-central Au+Au collisions at $\sqrt{s_{NN}}$=200 GeV.
Right Panel:
$R_{AA}$ of $\Upsilon$(1S+2S+3S) states as a function of the number of participants
in Au+Au collisions at $\sqrt{s_{NN}}$=200 GeV.
The solid black points are the $\Upsilon$ results, the red open points are the high $p_T>$ 5 GeV/c
$J/\psi$ results. 
The blue boxes at (0,1) represent the systematic and statistical
 uncertainty from the p+p cross section applying` to all shown $\Upsilon$ points,
while the grey boxes around the  $\Upsilon$ points are the systematic uncertainties from other
sources (see \cite{rosi_reed_qm2011} for details).
The red dotted line is the ratio of the total cross-section of $\Upsilon$(1S) over the  $\Upsilon$(1S+2S+3S).
The purple dashed line is the ratio of only the direct $\Upsilon$(1S) cross-section over the
total $\Upsilon$(1S+2S+3S).
}
\label{fig2}
\end{figure}


One way to quantify nuclear effects like the jet quenching is by using the
nuclear modification factor $R_{AA}$.
The $R_{AA}$ of fully reconstructed $D^0$ mesons, measured recently for the first time at RHIC,
 is shown in Figure~\ref{fig1} (left panel) to be consistent with 1, namely
with no suppression for $p_T<$ 3 GeV/c,  reflecting the binary scaling behaviour of the charm
production cross section
  \cite{yifei_zhang_qm2011}. 
The prediction of a Blast Wave fit using the freeze-out parameters of light-quark hadrons
(Figure~\ref{fig1} (left panel))
does not describe the $R_{AA}$ of $D^0$ mesons, which  may indicate that
  $D^0$ mesons freeze-out earlier than light-quark hadrons  \cite{yifei_zhang_qm2011}. 
In
 the high $p_T$ range, a suppression of electrons from open heavy flavour decays
appears prominently, in particular 
 the $R_{AA}$ of electrons from beauty as well as from 
charm meson decays at $p_T>5$ GeV/c have been measured to be both significantly suppressed
at the 90\% confidence level,
 in central
Au+Au collisions at 200 GeV
 \cite{bottom_RAA}.

Quarkonia can be suppressed by color screening in the plasma,
and different states have different dissociation temperatures.
The hierarchy of the suppression pattern of quarkonia
can therefore
   serve as a QGP signature and as a thermometer of the quark gluon plasma formed in these collisions.
Also, effects other than color screening
can contribute to the suppression of quarkonia, like e.g. suppression by hadronic comovers.
In addition, quarkonia can be regenerated from $q \overline{q}$ pairs 
counteracting a possible suppression.
 
\begin{figure}[htbp]
\begin{center}
\includegraphics[height=1.9in,width=2.23in]{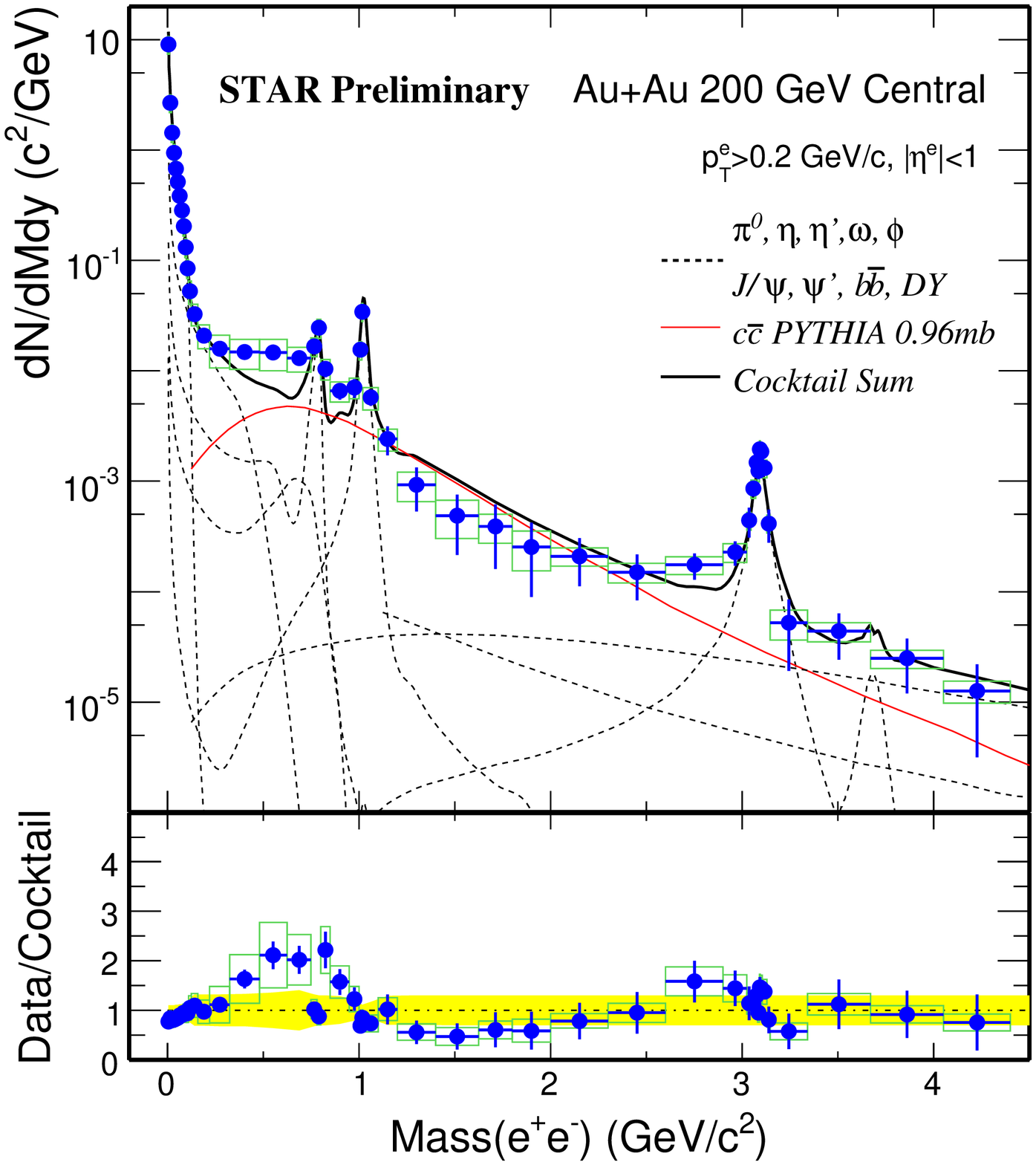}
\includegraphics[height=1.9in,width=2.23in]{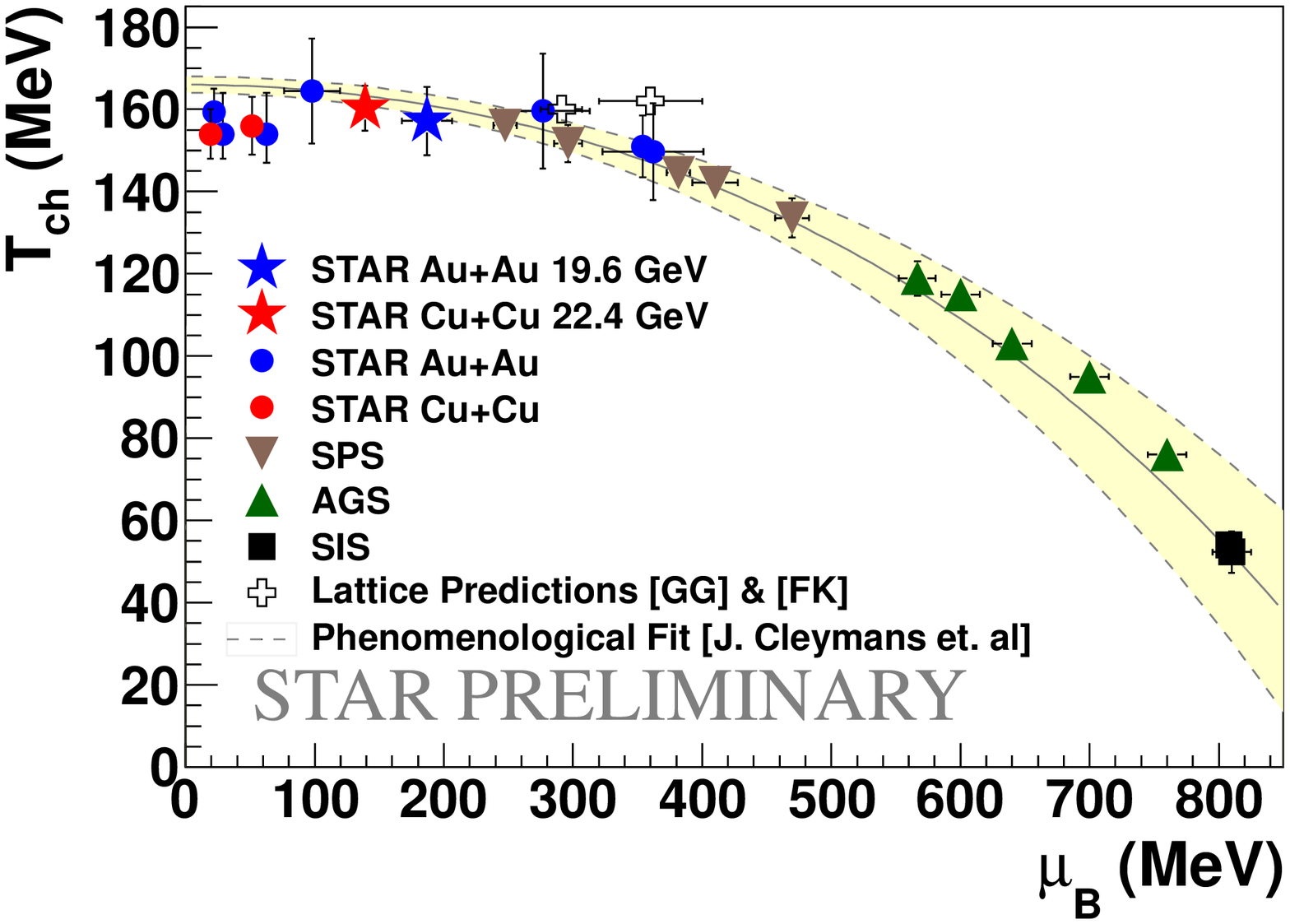}
\caption{
Left Panel: Invariant mass spectra $m(e^+e^-)$ from central
 Au+Au collisions at  $\sqrt{s_{NN}}$=200 GeV. 
The yellow band is the systematic error estimate on the cocktail.
Right Panel:
Temperature as a function of the baryochemical potential both at chemical freeze-out
extracted from data of several collision systems 
together with model predictions and fits.
} 
\label{fig4}
\end{center}
\end{figure}

The $J/\psi$ has been measured by STAR at midrapidity up to $p_T$ of ~9 GeV/c in
Au+Au collisions, and up to $p_T$ of ~14 GeV/c in $p+p$ collisions.
The $R_{AA}$ factor for the $J/\psi$ 
 has been found to be 
suppressed in central Au+Au collisions at $p_T>5$ GeV/c, 
while  it exhibits a larger suppression (a smaller $R_{AA}$)  at smaller $p_T$
(shown by the open circles measured by PHENIX)
(Figure~\ref{fig1} (right panel))
\cite{zebo_tang_qm2011}.
The $J/\psi$  measurement in peripheral Au+Au collisions 
is consistent with previous STAR measurement in Cu+Cu collisions at same $p_T$ 
and at same number of participant nucleons (Figure~\ref{fig1} (right panel) ) \cite{star_cucu_paper}.
The $J/\psi$ measured by STAR at $p_T>$ 5 GeV/c and in central Au+Au collisions
is less suppressed than the $J/\psi$  measured by CMS in Pb+Pb collisions at
$\sqrt{s}=$2.76 TeV at $p_T>$ 6.5 GeV/c and also at midrapidity
\cite{cms}.
This is consistent with a larger system size at the LHC.
However, cold nuclear matter effects have to be quantified with $d$+Au and $p$+Au collision data
as a function of the rapidity and $p_T$
to systematically understand the $J/\psi$ production in heavy ion collisions.
For a discussion of the $J/\psi$ suppression at low $p_T$ at RHIC and LHC please see \cite{xin_dong}.

Also a  different amount of $J/\psi$ regeneration at RHIC and LHC energies 
would  influence  the $J/\psi$ energy dependence.
STAR has measured the elliptic flow of $J/\psi$
(Figure~\ref{fig2} (left panel)) 
 to be consistent with zero 
in the $p_T$ range 2 to 8 GeV/c in mid-central Au+Au collisions at 200 GeV \cite{qiu_hao_sqm2011}.
This measurement disfavours $J/\psi$ production dominantly through coalescence from thermalized
$c$ and $\overline{c}$ at RHIC, assuming that charm quarks exhibit elliptic flow.

The $p_T$ dependence of $J/\psi$ in $p+p$ collisions at 200 GeV 
extended to low $p_T$ through the new TOF detector of STAR 
can constrain $J/\psi$ production models
and data indicate agreement with the Color Evaporation Model (CEM) at low $p_T$
\cite{leszek_kosarzewski_sqm2011}.
 New results on the $p_T$ dependence of the  $J/\psi$ polarization in 
$p+p$ collisions at 200 GeV are consistent with no polarization within uncertainties
\cite{barbara_trzeciak_sqm2011}.

Furthermore, STAR has measured the suppression of $\Upsilon$ states (1S+2S+3S) 
(Figure~\ref{fig2} (right panel))
in
central Au+Au collisions at 200 GeV for the first time at RHIC \cite{rosi_reed_qm2011}.
This suppression is consistent with suppression of the 2S and 3S $\Upsilon$ states
and only the 1S $\Upsilon$ state surviving. This hierarchy of suppression
 is expected within a color screening senario
due to the higher dissociation temperature
  $T_{dissociation} > 4 T_{critical}$
of the $\Upsilon$(1S) state \cite{satz}.
An indication of a  suppression of the  (2S+3S) $\Upsilon$ states
over the 1S $\Upsilon$ state
suppression  has been also observed in Pb+Pb over $p+p$ collisions at the LHC 
\cite{cms_bbbar}.

\section{Antimatter and dileptons}

STAR has reported recently in a publication in "Nature" the first observation of antihelium-4
 \cite{star_antihelium4}.
 The antihelium-4 yields measured
are important for background estimates of antimatter in space.
This result has been made possible with the help of the new TOF detector and a High Level Trigger (HLT)
in addition to the STAR TPC.

Further results of STAR using the TOF detector involve the measurements of the
dilepton invariant mass.
These measurements offer access to the in-medium modification of vector mesons
in the low mass region ($<$1.1 GeV) that could give informations related to the chiral symmetry restoration,
 while dileptons from thermal QGP radiation
and possible modification of correlated charm
can be studied in the intermediate mass region (1.1-3 GeV).
The dilepton  measurement in central Au+Au collisions 
(Figure~\ref{fig4} (left panel))
indicates modifications observed in the low and intermediate mass region, when compared to hadronic cocktail
simulations
\cite{bingchu_huang_sqm2011}.

New results on the $\phi$ meson indicate that the $\phi \rightarrow e^+ e^-$ result
is consistent with the previous $\phi \rightarrow K^+ K^-$ result, while no mass shift or
width broadening except for known detector resolution effects are observed
\cite{masayuki_wada_sqm2011}.

\begin{figure}[htb]
\begin{center}
\begin{tabular}{c}
\\
\includegraphics[height=1.9in,width=2.63in]{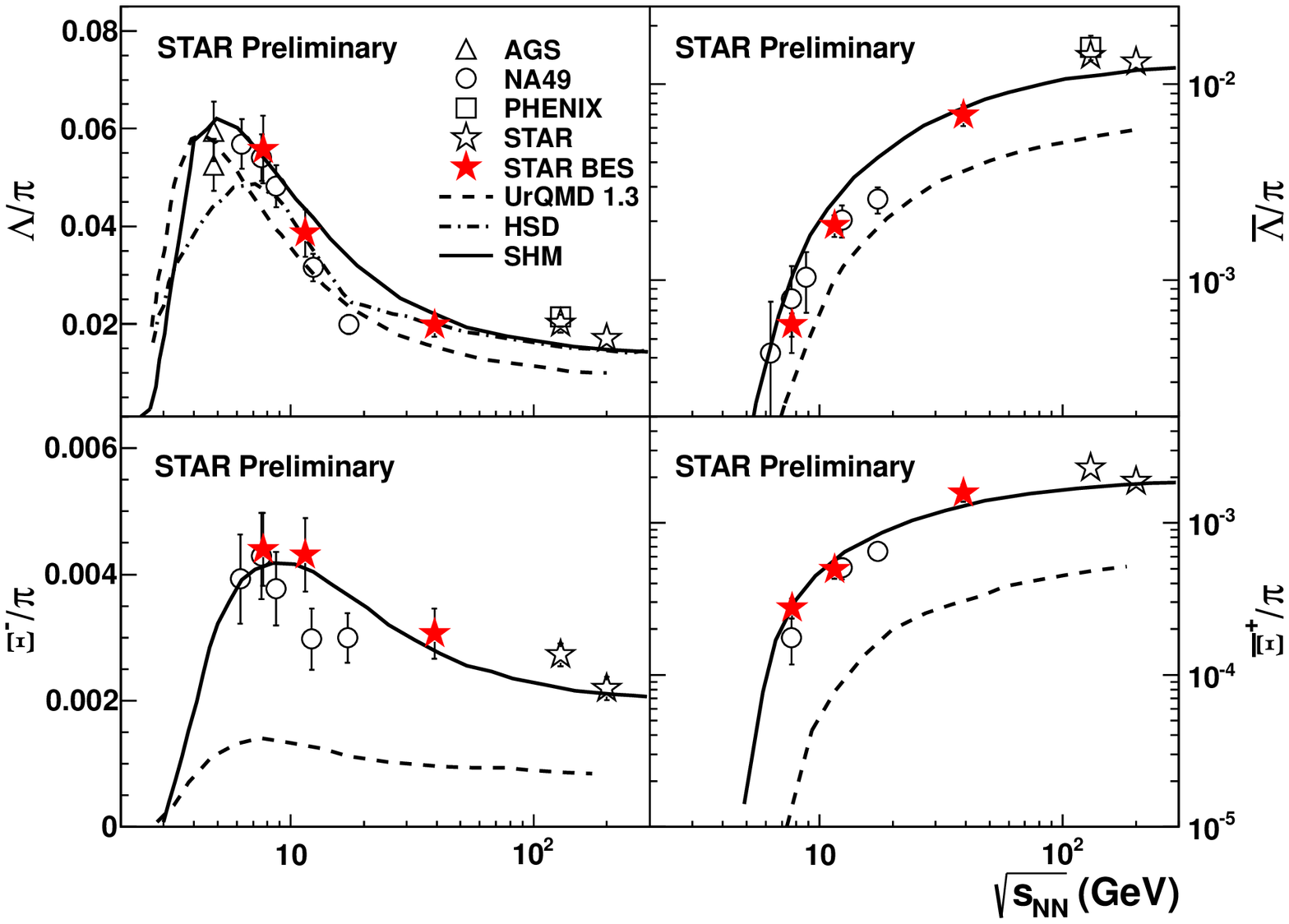}
\raisebox{-.02\height}
{
\includegraphics[height=1.9in,width=2.23in]{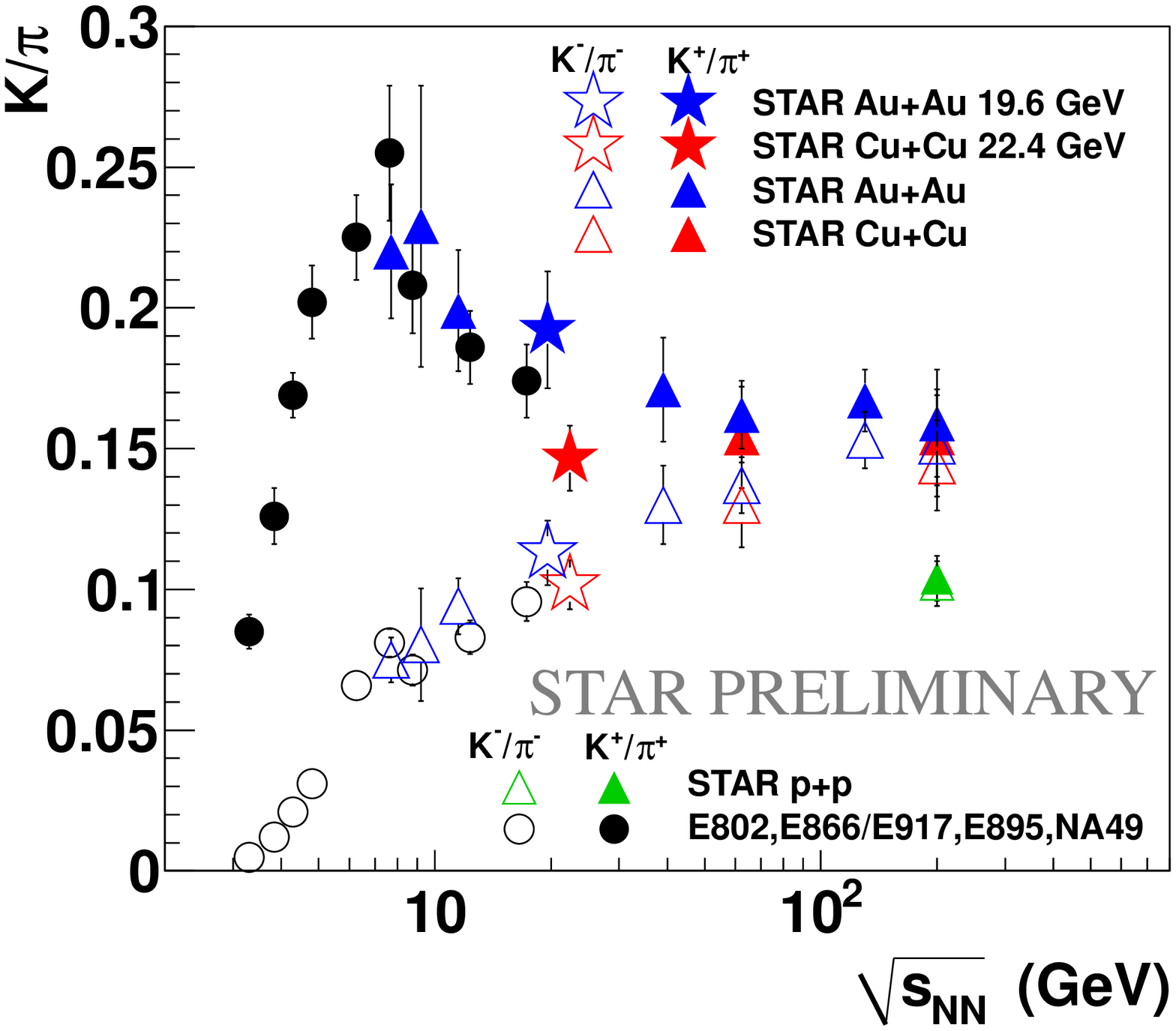}
}
\end{tabular}
\end{center}
\caption{
Left Panel: 
Ratios of 
 \lam , \alam, \xim\ and \axi to pions in central Au+Au and Pb+Pb collisions as a function of $\sqrt{s_{NN}}$
together with model predictions.
Right Panel:
 $K/\pi$ ratio vs. $\sqrt{s_{NN}}$ for the SIS, AGS, SPS and STAR data 
including the Beam Energy Scan.
} 
\label{fig5}
\end{figure}

\begin{figure}[!ht]
\begin{center}
\begin{tabular}{c}
\\
\includegraphics[width=4.63in]{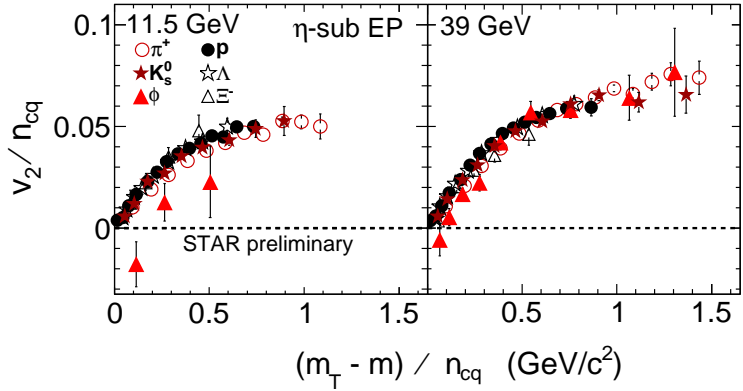}
\end{tabular}
\end{center}
\caption{
$v_2$ scaled by the number of constituent quarks ($ {\large  n_{cq}} $) as a function of $m_T-m$ over ${\large n_{cq} }$
for various particles in Au+Au collisions.
Please see \cite{shusu_shi_sqm2011} for more details.
}
\label{fig7}
\end{figure}

\section{Beam energy scan}

The goal of the "Beam Energy Scan" (BES) established at RHIC
is to measure with precision  the critical parameters
of the phase transition, and to discover a possible critical point 
and map out the QCD phase diagram \cite{BES}. 
The collision systems discussed here are 
Au+Au at $\sqrt{s_{NN}}$=7.7, 11.5, 19.6, 39, 62.4 and 200 GeV
and Cu+Cu at 22.4 GeV.
Figure \ref{fig4} (right panel)  shows 
the points measured by  STAR 
 representing the 
temperature ($T$(chem)) and baryochemical potential ($\mu_B$(chem)) at the time of chemical freeze-out
  \cite{orpheus_mall_sqm2011}.
  The $T$(chem)  and  $\mu_B$(chem)
 parameters have been obtained with a thermal model fit to particle ratios measured at each
energy.

A variety of strange particles, in particular
 \lam , \alam, \xim\ and \axi 
\cite{xianglei_zhu_sqm2011}
and kaons \cite{lokesh}
 have been reconstructed 
in Au+Au collisions at  $\sqrt{s_{NN}}$= 7.7 GeV, 11.5 GeV and 39 GeV
 using the STAR TPC.
Antibaryon to baryon ratios and yields  agree well with 
NA49 data at comparable energies.
 The  $\mu_B$(chem) in $A+A$ collisions has been found to decrease with increasing 
 collision energy, while the $T$(chem) is found to increase with increasing collision energy
 and saturate after the region around $\sqrt{s_{NN}}~$ 10 GeV \cite{lokesh}.
 The energy dependence of the strange particle ratios to pions
shown in Figure~\ref{fig5}
  reflects the above collision energy dependence of 
$T$(chem)  and  $\mu_B$(chem) \cite{lokesh},
  showing  maxima 
at low energies for the strange particles  produced also by 
 associated baryon-meson production e.g. $\Lambda K^+$
\cite{xianglei_zhu_sqm2011,lokesh}.
This is  expressed also by the agreement of 
the Statistical Hadron gas Model (SHM)
predictions \cite{SHM}
with the data across the whole energy range from AGS to top RHIC energies,
shown in Figure~\ref{fig5} (left panel).
This agreement indicates also the applicability of the grand canonical ensemble 
  assumed in this model,
to the chemical freeze-out conditions of A+A collisions in a large range of energies.

\begin{figure}[htb]
\begin{center}
\begin{tabular}{c}
\\
\includegraphics[width=2.33in]{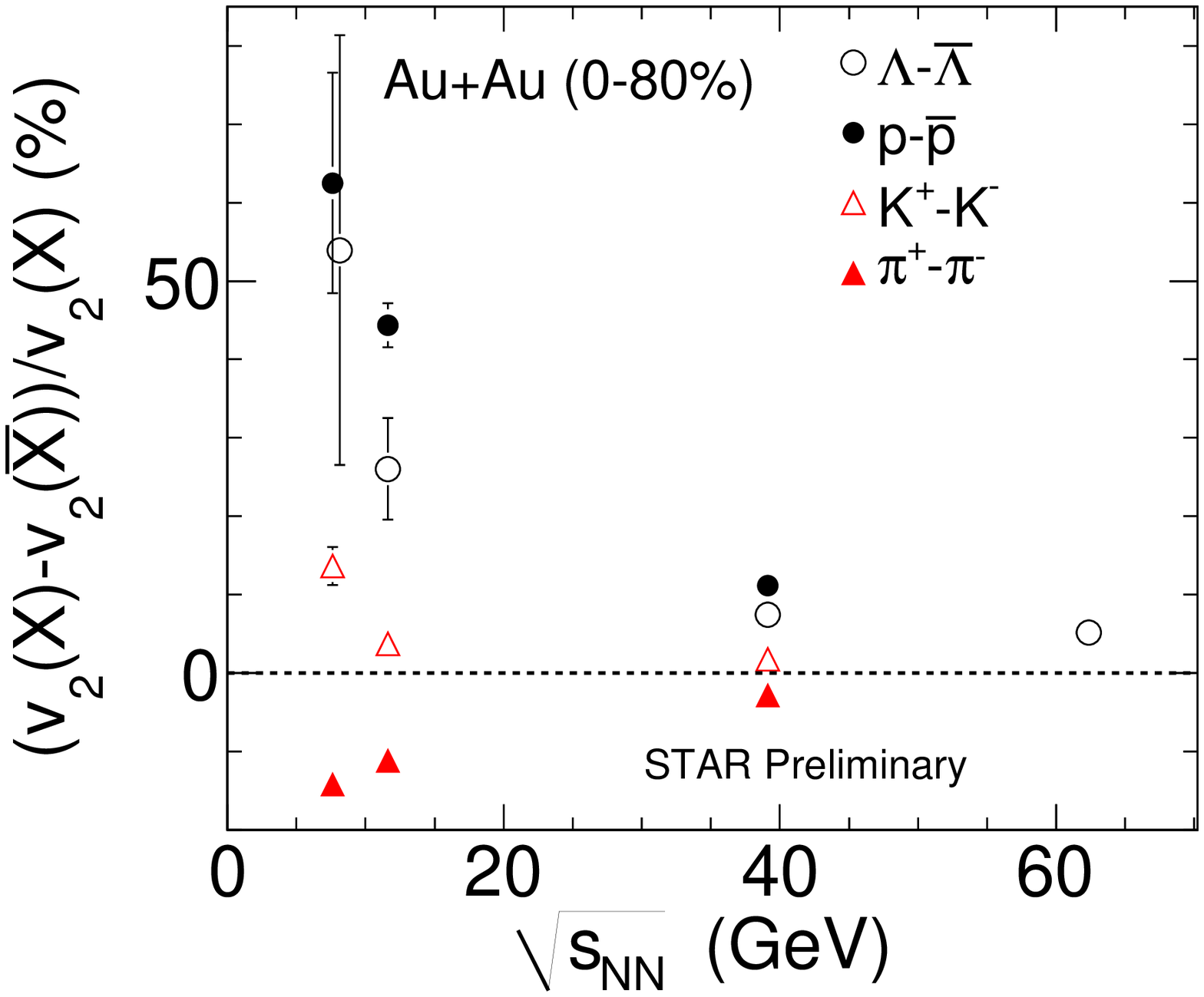}
\raisebox{-.2\height}
{
\includegraphics[width=2.33in]{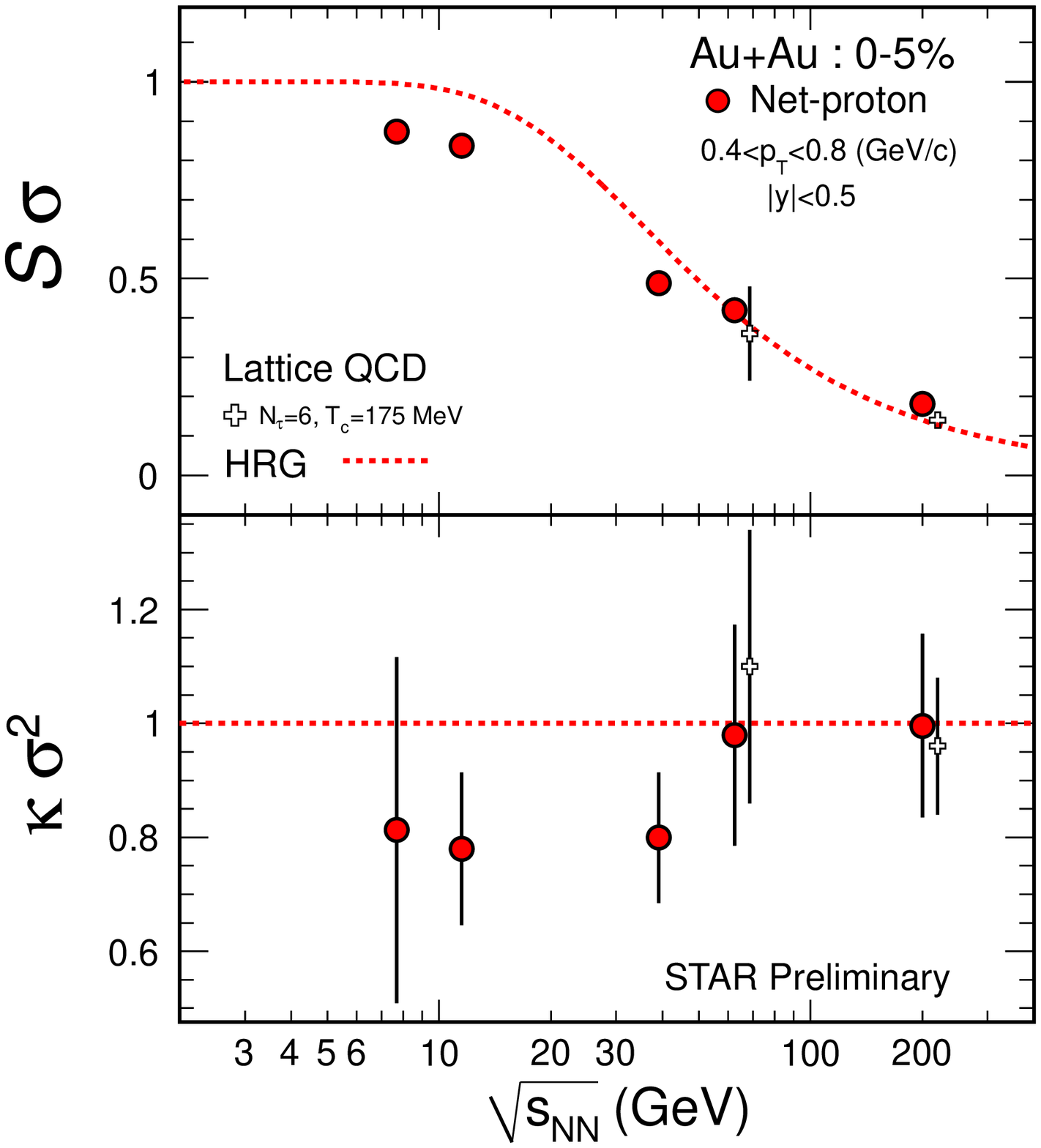}
}
\end{tabular}
\end{center}
\caption{
Left Panel:
The difference of $v_2$ for particles and antiparticles divided by the particle $v_2$ 
as a function of collision energy in Au+Au collisions.
Right Panel:
{Energy dependence of moment products ({\KV} and {\SD})
of net-proton distributions for 0-5\% most central Au+Au collisions.
The red dashed lines denote the HRG model calculations, and the
empty markers denote Lattice QCD results.}
} 
\label{fig6}
\end{figure}

The nuclear modification factor of the  $\phi$ meson in central with respect to peripheral
Au+Au collisions at $\sqrt{s_{NN}}$=39 GeV has been measured to show no significant suppression up to  $p_T$ 5 GeV/c
\cite{xiaoping_zhang_sqm2011}.
The elliptic flow coefficient $v_2$ of $\phi$ meson, measured up to $p_T$= 2 GeV/c at $\sqrt{s_{NN}}$=11.5 GeV,
 shows a deviation from the
$v_2$ of other hadrons in Au+Au collisions at $\sqrt{s_{NN}}$=11.5 GeV
(Figure~\ref{fig7})
\cite{md_nasim_sqm2011,shusu_shi_sqm2011}.

Furthermore,
the difference of $v_2$ between particles and antiparticles is found to be weakly dependent on energy from
the $\sqrt{s_{NN}}$=39 GeV on,  while it 
shows a significant deviation at energies lower than $\sqrt{s_{NN}}$=11.5 GeV
(Figure~\ref{fig6} (left panel)),
which increases with decreasing energy 
\cite{shusu_shi_sqm2011}.

Another interesting effect of a deviation at small energies is observed in the
proton and antiproton 
slope $ dv_1/dy' $  of the $v_1$ flow coefficient of the protons and antiprotons
at mid-rapidity as a function of beam energy at $\sqrt{s_{NN}}$=11.5 GeV
below which the proton slope changes sign, while the antiproton slope remains negative  
\cite{yadav_pandit_sqm2011}. 

Furthermore, STAR has measured dynamical fluctuations of several particle ratios on
an event-by-event basis in BES searching for a possible critical point.
The $K/\pi$ dynamical fluctuations from STAR show no significant
energy dependence in Au+Au collisions from $\sqrt{s_{NN}}$=7.7 to 200 GeV 
\cite{jian_tian_sqm2011,terence_tarnowski_sqm2011}.

Another remarkable result has been achieved by STAR while
 measuring the higher moments
 (variance $\sigma^2$, skewness S and kurtosis $\kappa$) 
of net-proton distributions in search for the QCD critical point.
In particular the moment products $\kappa$$\sigma^2$ and S$\sigma$ of net-proton distributions
in most central Au+Au collisions 
(Figure~\ref{fig6} (right panel))
are consistent with Lattice QCD and Hadron Resonance Gas (HRG)
 model calculations between
 $\sqrt{s_{NN}}$= 62.4 and 200 GeV, while the results are smaller than the HRG model estimates
at lower energies (at $\sqrt{s_{NN}}$=7.7, 11.5, 39 GeV)
\cite{xiaofeng_luo_sqm2011,netprotons}.

\section{Conclusions and outlook}

The STAR experiment at RHIC has entered a new era of high precision and high statistics
measurements with the help of recent major upgrades in particular the addition
of a barrel TOF detector, a Data Acquisition upgrade and a High Level Trigger.
The highly enhanced data statistics and the new identification capabilities 
lead to a number of new striking results.
STAR studies the properties of the new state of matter discovered at RHIC, the strongly interacting
Quark Gluon Plasma measuring
A+A collisions at the top RHIC energy and accessing low baryochemical potential of 
about 20 MeV.
Furthermore, STAR is exploring the QCD phase structure searching for
a possible critical point and the phase boundary
performing a beam energy scan with Au+Au collisions.
  New upgrades coming up in particular
 the silicon Heavy Flavour Tracker detector (HFT) and the
Muon Telescope Detector (MTD) heading to data taking in 2014 will allow for high
 precision measurements of  open heavy flavour, quarkonia and dimuon pairs
towards significant discoveries in the coming years.

\end{document}